# OPTIMAL DISCRETE-TIME OUTPUT FEEDBACK CONTROL FOR MULTI-AREA LOAD FREQUENCY CONTROL USING EVOLUTIONARY PROGRAMMING

A. Bensenouci[*] and A.M. Abdel Ghany[*]

[*]College of Technology at Al-Baha, Electrical Technology Department, Al-Baha, Saudi Arabia

*Abstract—* The interconnected power system presents a great challenge to both system analyzers and control designers. The load-frequency control (LFC) problem has gained much importance because of the complexity and size of modern interconnected power systems. In this work, the original (full) system is decomposed into subsystems using the overlapping decentralization technique. A discrete-time output feedback control is then designed using Evolutionary Programming (EP) technique. EP is selected since it is a good candidate for a global search for the optimum of a cost function that leads to the optimum output feedback controller gains in order to achieve the LFC requirements and improve its performance. The system performance is analyzed through simulating different disturbances and parameter variations over a wide range. Results from Dynamic Programming technique are also presented for completeness.

## I.  INTRODUCTION

The interconnected electrical power systems are complex multivariable large-scale dynamic systems in which system characteristics fluctuate with varying loads and varying generation schedules. Due to the complications involved, several challenges have been raised to achieve better operation and reliable performance.

The commonly used technique for LFC [1, 2] is based on the Area Control Error (ACE) that is a linear combination of power net-interchange and frequency errors. The main objective of the LFC requirements are to minimize the transient errors of the frequency and tie-line power, and to ensure zero steady-state errors of these quantities.

Several techniques and control strategies for LFC were proposed in the literature and were dealing with designing centralized and decentralized systems [3]. It is well known that the Implementation of a centralized controller poses certain difficulties when the number of interconnected areas increases. In addition, the LFC measurable signals such as frequency and tie-line power are in discrete form.  Moreover, the LFC problem has drawn wide attention using optimal control theory [4, 5]. Practical realization of the centralized LFC requires complete system state feedback measurements that might not be possible without the use of observers who may, in turn, complicate further the design and degrade the system performance. Therefore, it is essential to design alternative and more practical forms of discrete decentralized output feedback controller [6]. For this purpose, the original system (full) is decomposed into subsystems (areas), characterized by a stable operation when taken in isolation from each other, where local controllers are needed to make each area capable of handling its own load changes. Moreover, the centralized LFC might fail due either to the lack of information or computing capabilities beside the limitations on the information transferred by certain groups of sensors or actuators. A special attention should be given to the selected sampling interval and its impact on the system stability, computational effort needed, and the memory requirements. Adding to this, the effect of the changes in the feedback gains when the system is subjected to wide changes in the operating conditions.

Lately, several optimization techniques have been proposed and applied to LFC. The main focus of this work is to introduce an optimized output-feedback control (OFC) approach [6] for both centralized and decentralized LFC systems. The linear discrete OFC gains are derived using the Evolutionary Programming (EP) technique [7-10] that represents an excellent search tool for the optimum. The developed approach is applied to four-area power system connected in longitudinal form and where known measurements, in the control center, of the Area Control Error (ACE), are taken as output.

The superiority in implementing such controller resides in its simplicity, that is, no need for unmeasurable states or observers, and telemetry problems, and is suitable for on-line applications since only readily known or measured output and input historical data are used to predict the output vector.

The effectiveness of the procedure is demonstrated through diverse tests including a wide range of operating conditions namely, step and a ramp variations in power demand, and parameter changes.

## II. POWER SYSTEM MODELING AND DECOMPOSITION

### A. Continuous-time dynamic model

The continuous linear dynamic model, in state-space form, can be written as:

$$\begin{cases} \dot{x} = Ax + Bu \\ y = Cx + Du \end{cases} \quad (1)$$

Where
- $x$ state vector (nx1, n=18)
- $u$ control and disturbance vector (8x1) given by

$$u = [\Delta P_{d1} \Delta P_{c1} \Delta \Delta_{d2} \Delta P_{c2} \Delta P_{d2} \Delta P_{c3} \Delta P_{d4} \Delta P_{c4}]^t$$

$A$(nxn), $B$(nx8) and $C$(8xn) are constant matrices

The state variables and inputs are defined as follows:
- Incremental frequency deviations:
  $\Delta F_1 = x_1$, $\Delta F_2 = x_6$, $\Delta F_3 = x_{11}$, $\Delta F_4 = x_{15}$
- Incremental change in tie-line powers:
  $\Delta P_{tie1}=x_5$, $\Delta P_{tie2}=x_{10}-x_5$, $\Delta P_{tie3}=x_{14}-x_{10}$, $\Delta P_{tie4}=-x_{14}$
- Incremental load demand change (disturbances): $\Delta P_{di}$ (i=1-4)
- Incremental speed changer position (control inputs): $\Delta P_{ci}$ (i=1-4)

### B. System Decomposition

The overlapping decentralization technique proposed by Siljak [11] is employed to decompose the full system into subsystems (areas). In this technique, the areas represent the subsystems while the change in tie-line power is the overlapping part. The system is decomposed into four components $\tilde{x}_1 = [x_1\ x_2\ x_3\ x_4\ x_5]^t$, $\tilde{x}_2 = [x_5\ x_6\ x_7\ x_8\ x_9\ x_{10}]^t$, $\tilde{x}_3 = [x_{10}\ x_{11}\ x_{12}\ x_{13}\ x_{14}]^t$ and $\tilde{x}_4 = [x_{14}\ x_{15}\ x_{16}\ x_{17}\ x_{18}]^t$ representing the 4 areas.

With this representation, the system becomes

$$\begin{bmatrix} \tilde{x}_1 \\ \tilde{x}_2 \\ \tilde{x}_3 \\ \tilde{x}_4 \end{bmatrix} = \begin{bmatrix} A_{11} & \cdots & A_{14} \\ \cdots & \cdots & \cdots \\ A_{41} & \cdots & A_{44} \end{bmatrix} \begin{bmatrix} \tilde{x}_1 \\ \tilde{x}_1 \\ \tilde{x}_1 \\ \tilde{x}_1 \end{bmatrix} + \begin{bmatrix} B_{11} & \cdots & B_{14} \\ \cdots & \cdots & \cdots \\ B_{41} & \cdots & B_{44} \end{bmatrix} \begin{bmatrix} u_1 \\ u_2 \\ u_3 \\ u_4 \end{bmatrix} \quad (2)$$

Where $A_{ij}$, $B_{ij}$ (i, j=1-3) are subsystem matrices whose elements depend on the system parameters and

$$u_i = [\Delta P_{di}, \Delta P_{ci}]^t \quad (i=1-4) \quad (3)$$

The new vector $\tilde{x}$ is related to $x$ by:
$$\tilde{x} = Tx. \quad (4)$$

The expanded system can be reformulated using overlapping subsystems as follows:

$$\dot{\tilde{x}} = \tilde{A}\tilde{x} + \tilde{B}u \quad (5)$$

where

$$\tilde{A} = TAT^* + M$$
$$\tilde{B} = TB$$
$$T^* = (T^t T)^{-1} T^t$$

Using the previous equations, the expanded system can be described as:

$$\dot{\tilde{x}}_i = \tilde{A}_i \tilde{x}_i + \tilde{B}_i u_i + \sum_{i \neq j} \tilde{A}_{ij} \tilde{x}_j + \sum_{i \neq j} \tilde{B}_{ij} u_j \quad (6)$$

$i = 1,..,4 \quad j = 1,..4$

where $\tilde{A}_i, \tilde{B}_i$ (i = 1 - 4) are the matrices corresponding to the four decoupled subsystems. The control input to each subsystem is defined by $u_i = [\Delta P_{di}, \Delta P_{ci}]^t$. For the control purpose, assume weak coupling element such as $\tilde{A}_{ij}$ & $\tilde{B}_{ij}$ ($i \neq j$) can be neglected. Therefore, the decoupled controlled subsystems are given by:

$$\dot{\tilde{x}}_i = \tilde{A}_i \tilde{x}_i + \tilde{B}_i u_i \quad (i = 1,..,4) \quad (7)$$

### C. Discrete-time dynamic model

A discrete-time model [7] for each subsystem can be obtained from eqn (7) as

$$\begin{cases} x_{k+1} = \Phi x_k + \Delta u_k \\ y_{k+1} = Cx_k + Du_k \end{cases} \quad (8)$$

where, $x_k = x(kT_S)$, $u_k = u(kT_S)$ are specified at $kT_S$, $k=0,1,...$ and $\Phi$, $\Delta$ are the state transition and input driving matrices, respectively.

## III. DISCRETE-TIME OUTPUT FEEDBACK

The state prediction equation of the discrete-time linear model described by (8) can take the form:

$$x_{k+1} = F_5 w_k + F_4 u_k \quad (9)$$

The output-prediction equation has the form:

$$y_{k+1} = \alpha z_k + \beta v_k \quad (10)$$

The prediction equation of the augmented vector $w_k$ is

$$w_{k+1} = \theta w_k + \Omega u_k \quad (11)$$

where

$$z_k = [y_k \; y_{k-1} \; \cdots \; y_{k-N+1}]^t$$

$$v_k = [u_k \; u_{k-1} \; \cdots \; u_{k-N+1}]^t$$

$$w_k = [z_k | v_k]^t$$

The matrices $\alpha$, $\beta$, $\theta$, $\Omega$ are defined in [6]. N is the measurement number of output and input from $t=kT_s$ back to $t=(k-N+1)T_s$. The minimum number of previous measurement vectors N is selected such that $N \geq n/p$ where p is the number of outputs.

Equation (10) completely defines the process dynamics without reference to the state vector **x**.

A state feedback optimal control law $u_k = F_s w_k$ is determined from the minimization of the quadratic-performance index of the form:

$$J_S = \sum_{k=0}^{r} \left[ x_{k+1}^t Q_s x_{k+1} + u_k^t H_s u_k \right] \quad (12)$$

Similarly, an output feedback optimal control law $u_k = F_o w_k$ is determined from the minimization of the quadratic-performance index of the form:

$$J = \sum_{k=0}^{r} \left[ w_{k+1}^t Q_o w_{k+1} + u_k^t H_o u_k \right] \quad (13)$$

The two performance indexes given by (12) and (13) are equivalent if (9) is substituted into (12) to get

$$J_O = \sum_{k=0}^{r} \left[ w_k^t Q w_k + u_k^t R w_k + u_k^t S u_k \right] \quad (14)$$

where

$$\begin{cases} Q = F_5^t Q_s F_5 \\ R = 2 F_4^t Q_s F_5 \\ S = F_4^t Q_s F_4 + H_s \end{cases}$$

To obtain a constant control matrix $F_o$ for the output feedback, substitute $u_k = F_o w_k$ into (14) to get

$$J_O = \sum_{k=0}^{r} \left[ w_k^t G w_k \right] \text{ with } G = Q + F_o^t R + F_o^t R \quad (15)$$

It is worth noting that in order to reach the global optimum of $J_0$, the weight matrices $H_s$ and $Q_s$ are assumed to be symmetric positive definite matrices as shown in [4]. Besides, when using the output feedback representation, the closed loop eigenvalues can be determined from (11) and $u_k = F_o w_k$ to get $w_{k+1} = [\theta - \Omega F_o] w_k$. The eigenvalues of $A_{CL} = [\Phi - \Delta F_s]$ depend on the gain values of the output feedback matrix $F_o$ whose values are bounded in a stable interval that is, $F_{o,min} \leq F_{o,ij} \leq F_{o,max}$, for EP use. The system response converges to the global optimum.

## IV. OPTIMIZATION PROBLEM FORMULATION

On the basis of assumed sampling time interval $T_s$, the optimization problem is thus defined as:

Find $F_0$ that minimizes $J_O = \sum_{k=0}^{r} \left[ w_k^t G w_k \right]$ with respect to $u_k = F_o w_k$ where $G = Q + F_o^t R + F_o^t R F_o$.

To calculate the output feedback control gains $F_o$, two optimization solver techniques are presented and applied to LFC, namely, the Evolutionary Programming (EP) and the Dynamic Programming (DP) techniques.

## V. EVOLUTIONARY PROGRAMMING

Evolutionary Programming (EP) Technique with its capability to solve complex optimization problems represents a very attractive solution tool for many industrial applications. Unlike conventional optimization techniques, the EP starts with little or no knowledge of the correct solution, therefore, EP works with a population of points that represents different potential solutions. For each generation, all of the population points are evaluated based on a certain objective function. The fittest points have more chances of evolving to the next generation. Therefore, EP can eliminate and overcome the drawbacks found in conventional optimization techniques.

The advantages of EP can be summarized as follows:
1- It can easily deal with non-smooth, non-continuous and non-differentiable objective functions that are close to real-life problems.
2- It uses probabilistic translation rules to make decisions. Hence, EP is a kind of stochastic optimization algorithm that can search a complicated and uncertain space to find the global or near- global optimal solution. This makes EP more flexible and robust than conventional methods.

In the EP algorithm, the iterative application can be described in the five main steps;
1. reproduction (initialization)
2. mutation,
3. statistics

4. tournament
5. selection and stopping criteria

## VI. DYNAMIC PROGRAMMING

Dynamic Programming (DP) technique is used to minimize $J_o$ at several stages starting from initial stage k=0 and moving backward until stage k=r. If r is large enough, the DP algorithm converges to $\mathbf{F}_o$ that is constant. The multi-stage dynamic programming algorithm is [4] summarized as follows:

Step1: Initialization process
   $\sigma = 0$
   Compute
   $\eta = R + 2*\Omega^t*\sigma*\theta$
   $\mu = S + \Omega^t \sigma \Omega$
   $F = -0.5*\mu^{-1}\eta$
   k=1
Step 2: Iterate while k>0 & |F-F$_0$| > tolerance, do
   { $F_0 = F$
   $\sigma = Q + \theta^t \sigma \theta + F^t \eta + F^t \mu F$
   $\mu = S + \Omega^t \sigma \Omega$
   $\eta = R + 2* \Omega^t*\sigma*\theta$
   $F = -0.5*\mu^{-1}\eta$
   k = k+1;
   }
   $F_D = F$

Where $\mathbf{F}_o = \mathbf{F}_D$ is the Dynamic programming matrix gains.

## VII. SIMULATION RESULTS

The simulation is done using MATLAB Platform. To make it more realistic, the linear continuous system is converted to a discrete-time one using zero-order hold with a specific sampling time. The LFC system comprises 4 areas; one hydro and 3 steam with and without reheat, as shown in Fig. 1. Both centralized and decentralized LFCs were considered. The discrete-time output feedback gains were optimized using Evolutionary Programming (EP) and Dynamic Programming (DP) techniques. Four tests were considered for verification of the effectiveness of the proposed controllers.

*Test 1*: Base Case Operating condition

*A. Centralized LFC with EP*

The output feedback is designed for the original or full system as one block. The design problem was to find $\mathbf{F}_c$ such that $\mathbf{u}=\mathbf{F}_c*\mathbf{w}$, where $\mathbf{F}_c$, $\mathbf{w}$ and $\mathbf{u}$ are vectors of dimension, respectively, (4x36), (36x1) and (4x1).

The dynamic response of the centralized LFC was very similar to the decentralized ones, so the only Fig.s presented are those of the decentralized case.

*B. Decentralized LFC with EP*

The output feedback is designed for the full system after processing a decoupling between the 4 areas using Siljak technique. The design problem was to find $\mathbf{F}_d$ such that $u_i=\mathbf{F}_{di}*\mathbf{w}_i$ (i=1-4). The evolutionary technique (EP) is used for this purpose. The vectors $\mathbf{F}_{di}$, $\mathbf{w}_i$ and $u_i$ are of dimension, respectively, (1x9), (9x1) and (1x1).

Fig. 2 shows the dynamic response of the area control errors $ACE_i$ and the area frequencies $\Delta F_i$ (i=1-4) when a step increase of 1% in the power demand $\Delta P_{di}$ (i=1-4) is applied to each area of the system. The area control errors and the tie-lines are not included but recover to zero steady state error after some time delay from the application of the disturbance.

Both centralized and decentralized LFCs give similar responses. But, since the decentralized one has other positive practical features over the centralized one, as mentioned previously, the former (decentralized) is used in subsequent tests.

*C. Decentralized LFC with DP*

The output feedback is designed using Dynamic Programming for each of the 4 areas of the system. The design problem was to find $\mathbf{F}_D$ such that $u_i=\mathbf{F}_{Di}*\mathbf{w}_i$ (i=1-4). The vectors $\mathbf{F}_{Di}$, $\mathbf{w}_i$ and $u_i$ (i=1-4) are of dimension (1x9), (9x1) and (1x1), respectively. The multi-stage Dynamic Programming Technique is used for completeness.

Fig. 3 shows the dynamic response of the area frequencies $\Delta F_i$ (i=1-4) when a step increase of 1% in the power demand $\Delta P_{di}$ (i=1-4) is applied to each area of the system. The system is shown to recover to nominal values of zero in all area frequencies.

*Test 2*: Wide Parameter Variation(Decentralized & EP)

In this test, 50% increase in generator time constant $t_{pi}$, speed droop $r_i$, tie-line between area "i" and area "j" parameter $t_{ij}$, area control gain $B_i$ (i, j=1-4). Fig.s 4 shows the dynamic response of $\Delta F_i$ (i=1-4) when a step in power demand of $\Delta P_{di}=1\%$ is applied to each area. The system is show to recover to nominal values of zero after some delay.

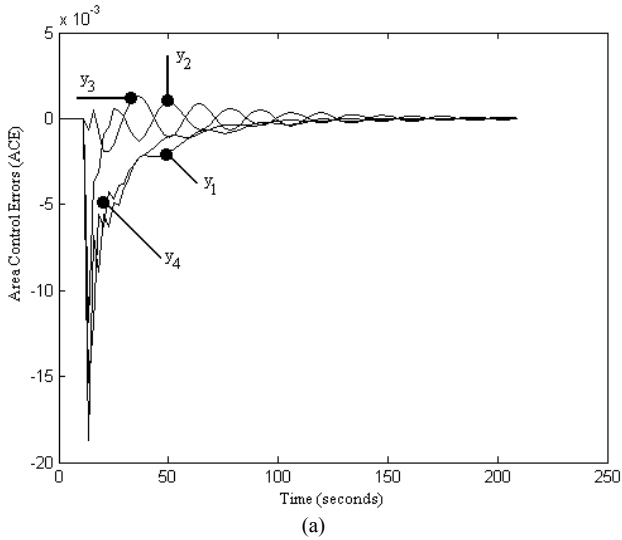

(a)

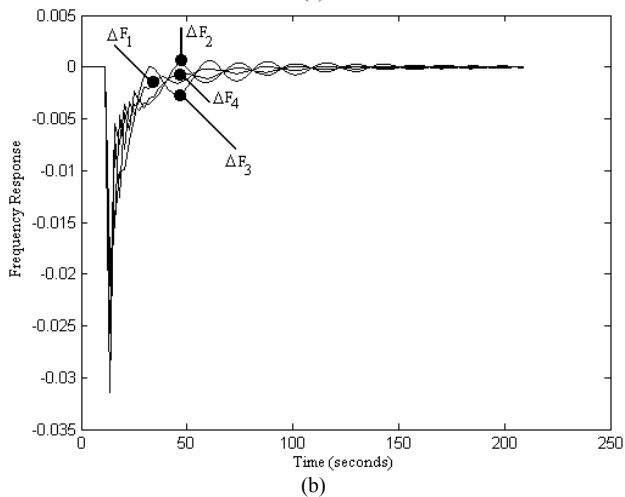

(b)

Fig. 2. Frequency Response due to $\Delta P_{di}$=+1% (i=1-4) in all areas (Test 1: Decentralized LFC & EP)

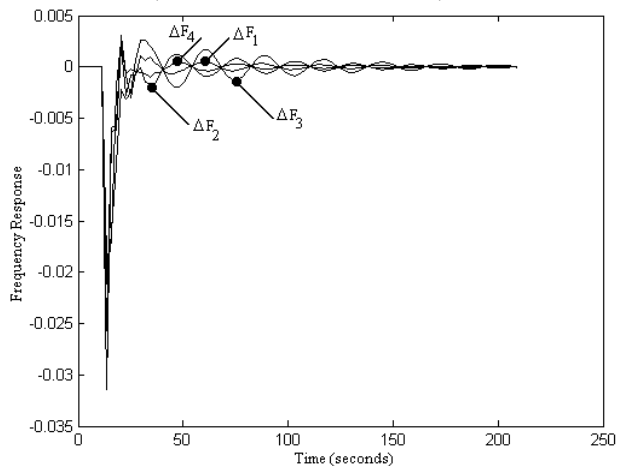

Fig. 3. Frequency Response due to $\Delta P_{di}$=+1% (i=1-4) in all areas (Test 1: Decentralized LFC & DP)

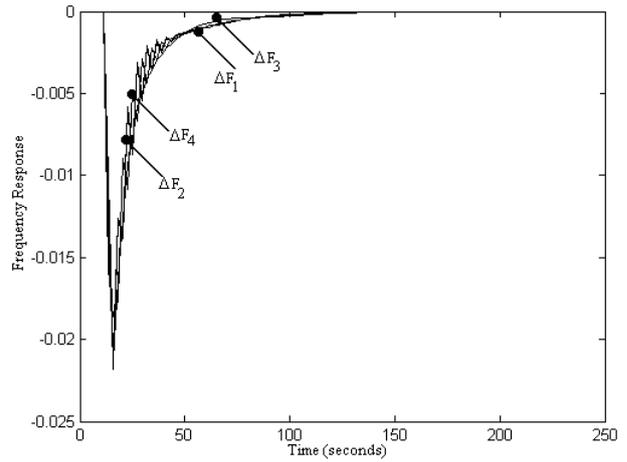

Fig. 4. Frequency Response due to an increase of 50% in $t_{pi}$, $r_i$, $B_i$, $t_{ij}$ (i,j=1-4) from their nominal values (Test 2: Decentralized LFC & EP)

*Test 3: Disturbance Variations (Decentralized & EP)*

Fig. 5 shows the dynamic response of $\Delta F_i$ (i=1-4) following a variation in each one of the area power demand $\Delta P_{di}$ (i=1-4) as seen in Fig. 5 (b). This variation covers both tracking (ramp) and regulation (step change). It is clear that the system kept its stability and responds effectively within an acceptable period of time before arriving to zero steady-state.

## VIII. CONCLUSION

This paper presented the design of an optimal discrete-time centralized and decentralized LFC to achieve improvements in transient and steady state responses. For the decentralized LFC case, the overlapping decentralization technique proposed by Siljak is used to decompose power systems into areas. The optimization technique used is the Evolutionary Programming (EP) and Dynamic Programming (DP) techniques. The presented design steps are based on the state-space representation of the system.

EP is selected as a candidate to search for the optimal gain settings of the output feedback controller. The proposed approach has been applied to a four-area power system where different disturbances and parameter changes were applied. The results show that the proposed controller is effective and recovers to normal operation of zero steady state errors following the mentioned perturbations. EP can be effectively applied to more complex large-scale nonlinear systems.

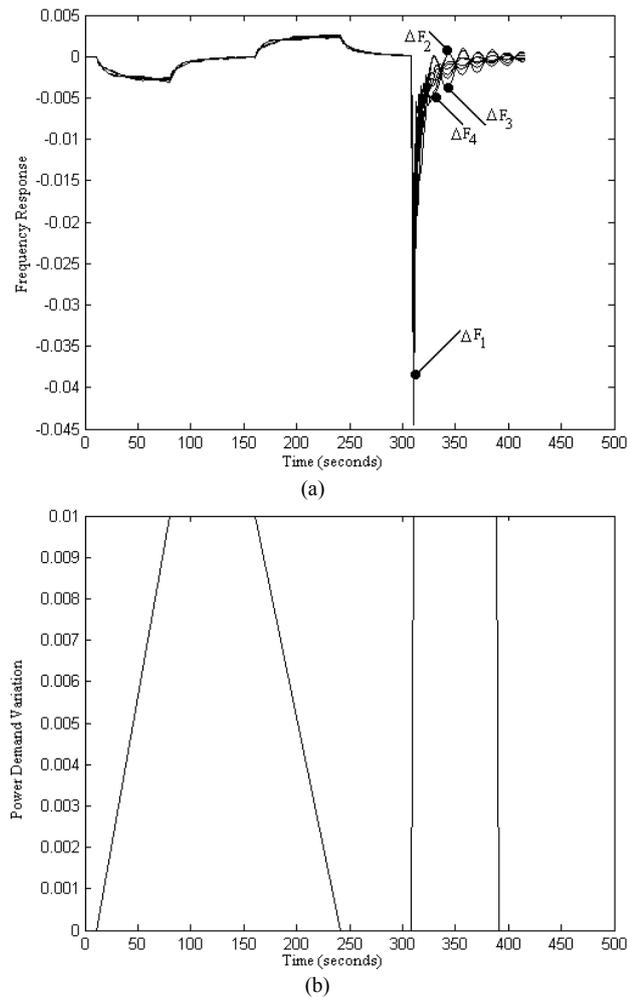

Fig. 5. System Response due to $\Delta P_{di}$ variation
(Test 3: Decentralized LFC & EP)

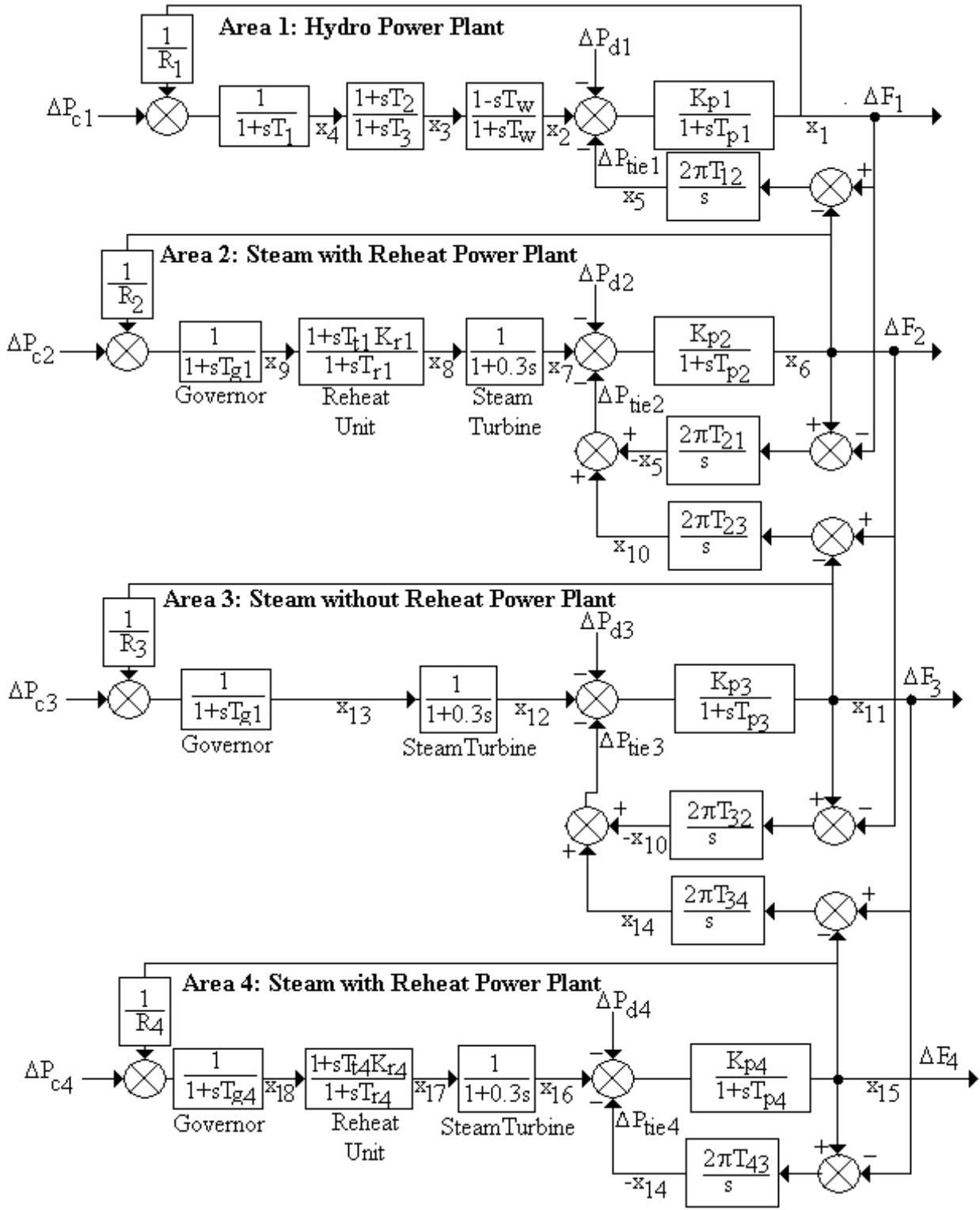

Fig. 1. Four areas interconnected Power System